\documentclass[aps,pre,twocolumn,superscriptaddress,showpacs]{revtex4}
\usepackage{dcolumn}
\usepackage{graphicx}
\usepackage[dvips]{hyperref}
\usepackage[TS1,OT1,T1]{fontenc}

\begin{document}

 \newtheorem{thm}{Theorm}
 \newtheorem{defn}{Definition}
 \newtheorem{cor}{collory}
 \newtheorem{lem}{lemma}
 \newtheorem{prop}{proposition}


\title{A weighted network evolution model based on passenger behavior}



\author{Yihong Hu}
\email{051025007@fudan.edu.cn}
 \affiliation{Department of Management
 Science, Fudan University, Shanghai 200433, China}

\author{Daoli Zhu}
\affiliation{Department of Management Science, Fudan University,
Shanghai 200433, China} \affiliation{Institute of Shanghai logistics
development, Shanghai 200433, China}

\author{Nianqu Zhu}
\affiliation{Department of Management Science, Fudan University,
Shanghai 200433, China}


\date{\today}

\begin{abstract}
This paper presents an evolution model of weighted networks in which
the structural growth and weight dynamics are driven by human
behavior, i.e. passenger route choice behavior. Transportation
networks grow due to people's increasing travel demand and the
pattern of growth is determined by their route choice behavior. In
airline networks passengers often transfer from a third airport
instead of flying directly to the destination, which contributes to
the hubs formation and finally the scale-free statistical property.
In this model we assume at each time step there emerges a new node
with $m$ travel destinations. Then the new node either connects
destination directly with the probability $p$ or transfers from a
third node with the probability $1-p$. The analytical result shows
degree and strength both obey power-law distribution with the
exponent between 2.33 and 3 depending on $p$. The weights also obey
power-law distribution. The clustering coefficient, degree
assortatively coefficient and degree-strength correlation are all
dependent on the probability $p$. This model can also be used in
social networks.
\end{abstract}



\maketitle

Most networks in real world are weighted. Recently weighted networks
dynamics have attracted much attention. People have developed many
new models to understand weighted networks evolution mechanism based
on the Barab\'{a}si-Albert (BA) model \cite{1} which first proposed
preferential attachment for networks evolution. Antal-Krapivsky (AK)
model \cite{2} relaxes degree preferential attachment to weight
driven attachment mechanism. Barrat-Barth\'{e}l\'{e}my-Vespignani
(BBV) model \cite{3,4} adds local weight reinforcement dynamics to
the structure growth and weight-driven mechanism. Dorogovtsev-Mendes
(DM) model gets similar result with BBV model. W. Wang et al\cite{6}
gives a global weight reinforcement mechanism to structure growth
and weight driven attachment to model traffic networks evolution.
Guimera and Amaral \cite{7} models world-wide airport networks
including spatial information as weight. Researchers begin to put
more focus on weight dynamics and weight interaction with other
factors \cite{Boccalettia}.

The above papers all assume the weight changes automatically at each
step due to weight and topology coupling mechanism. However this
assumption is too rough to understand underlying driving factors and
helps little to control and manipulate networks. Transportation
networks are good examples to explain this argument. Traffic is
composed by passengers who are decision makers. They decide which
routes to choose and then changes the traffic on routes. The
shortest path finding behavior and congestion phenomenon in networks
have been studied by many papers \cite{B, Wu, Danila, Danila2,
Douglas}. Passengers collective route selection behavior play a
critical role in the hubs formation and the pattern of network
growth. For example, in airline industry airline companies provide
low price for those who are willing to take transfer flights.
Passengers make their decisions based on own time and money
consideration. Some sensitive to price select transfer flights and
contribute to traffic in hubs. If all passengers are not sensitive
and don't want to transfer, the hubs and thus the popular
hub-and-spoke operation model in airline industry since 1970's
cannot come into being. And the world-wide airline networks would
not display the structure nowadays. So here we propose a new
approach to understand transportation network evolution taking
passenger's route selection behavior into consideration in addition
to weight and topology coupling mechanism.

The model assumes at each stage there emerges a new node with $m$
destinations, i.e. new travel demand is created. A passenger who
wants to go from origin to destination select his route. He can
either fly directly to the destination or transfer at a third place.
The probability of taking non-stop flight is $p$ and that of
transfer flight is $1-p$. In the transfer case, the passenger
determines transfer node based on the weight between transfer node
and destination. Analytical results show degree and strength both
obey power-law distribution with the exponent between $2.33$ and $3$
depending on $p$. The weights also obey power-law distribution. If
$p=1$, the model recover the strength-driven AK model. If we let
$1-p=\delta$, it reproduces the results of the BBV model. Simulation
confirms the theoretical results. This model is not limited in
explaining transportation networks. It can also used to explain
social networks which we will discuss later.

The detailed model is described as follows. The initial network has
$N_0$ nodes with existing links. There are initial weights $w_0$ on
the links. The networks evolutes according to the following rules:

\begin{itemize}
 \item {\it Growth}. In each step a new node $n$ emerges with $m$ destinations.
The probability of one old node $i$ to be the destination is
determined by the ratio of its strength $S_i$ to the total strength
of all nodes:

\begin{equation}
\pi_{i}=\frac{S_i}{\sum_l S_l}
\end{equation}

which suggests that nodes with large strength have large possibility
to be travel destinations.

\item {\it Passenger's route choice}. Then the new node $n$ either connects node $i$ with probability
$p$ or transfers at a third node $j$ with probability $1-p$. In the
former case, a new edge with initial weight $w_0$ between $n$ and
$i$ is established. In the latter case, things are little
complicated. Passengers have to determine which node to choose as
the transfer node. They only consider destination's neighbors which
already have links with the destination. The probability of node
$j\epsilon \nu(i)$ being chosen as the transfer node is:

\begin{equation}
\pi_{n-j}=\frac{w_{ij}}{S_i}
\end{equation}

Here $\nu(i)$ is the neighbors set of node $i$. $w_{ij}$ is the
weight on edge between vertex $i$ and $j$ representing the traffic.
Once node $j$ is chosen, a new edge with initial weight $w_0$
between vertex $n$ and $j$ is created, and traffic between vertex
$i$ and $j$ increases by $w_0$. The quantity $w_0$ set the scale of
weight. We can set $w_0=1$ without impact to the model.
\end{itemize}

With the above two rules, the network grows and stops when nodes
reach a preset large number $N$.

The rules has practical implications. In the first rule, a great
number of traffic throughput indicate the city is more developed and
more attractive to people than those city having little traffic.
This rule is identical to those strength-driven models. In the
second rule, there are two approaches to get to the destination. If
passengers choose transfer mode, they prefer those nodes having
frequent flights and large traffic between the destination and
transfer node because frequent and large traffic can bring
convenience, low price and less waiting time.

\begin{figure}[th]
\begin{center}
\setlength{\unitlength}{1mm}
\begin{picture}(60,30)
\linethickness{2pt} \thicklines \qbezier[20](0,0)(0,10)(0,22)
\thicklines \qbezier(0,22)(9.525,16.5)(19.05,11)
\qbezier(0,0)(9.525,5.5)(19.05,11)\put(0,0){\circle*{2}}
\put(0,22){\circle*{2}} \put(59.05,11){\circle*{2}}
\put(40,0){\line(0,1){22}} \thicklines
\qbezier[20](40,0)(49.525,5.5)(59.05,11)
\qbezier[20](40,22)(49.525,16.5)(59.05,11) \put(40,0){\circle*{2}}
\put(40,22){\circle*{2}} \put(19.05,11){\circle*{2}}
\put(21,11){$j$} \put(61,11){$j$} \put(0,24){$n$} \put(40,24){$n$}
\put(0,-4){$i$} \put(40,-4){$i$} \put(30,-10){\footnotesize{(b)
connect directly with $p$}} \put(-10,-10){\footnotesize{(a) transfer
with $1-p$}}
\end{picture}
\end{center}
\vspace*{20pt} \caption{Visual description of evolution mechanism.
New node $n$ emerges at each step. The old node $i$ becomes travel
destination with probability $S_i/\sum_l S_l$. (a) Passengers select
one neighbor $j$ of node $i$, to transfer with probability
$w_{ij}/{S_i}$. The weights update as $w_{ij}=w_{ij}+1$ and
$w_{nj}=1$. (b) Passenger go to node $i$ directly creating a new
link between $n$ and $i$. The weight $w_{ni}=1$. \label{f1}}
\end{figure}
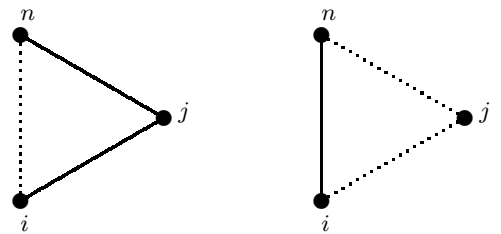

This model is also applicable to social networks. For example, in
author networks a new researcher is attracted to some famous
scientist, but he has no approach to know him directly. Instead he
finds that it's better to know the scientist's acquaintance with
plenty of connections or cooperations between them. With the
introduction of that acquaintance, he can finally get to know the
scientist. This will also happen in friends networks, actor
networks, etc.

To analyze the distribution of degree and strength, we notice the
model time is measured by the number of nodes added to the network,
i.e., $t=N-N_0$. And  the natural time scale of the model dynamics
is the network size $N$. We treat strength $S$, degree $K$, time $t$
as continuous variables \cite{9}. When a new node $n$ is added into
the network, the strength $S_i$ of an already present vertex $i$ can
increase by 1 if it is selected as a destination or increase by 2 if
it is selected as a transfer node. And the degree $K_i$ of the
vertex $i$ will only increase by $1$ whether it is chosen as
destination or transfer node. We have the following equations:

\begin{equation}
\frac{dS_i}{dt}=m\frac{S_i}{\sum_l S_l}+(1-p)\sum_{j\epsilon \nu(i)}
2m\frac{S_i}{\sum_l S_l}\frac{w_{ij}}{S_j}\label{strengthequation}
\end{equation}

\begin{equation}
\frac{dK_i}{dt}=m\frac{S_i}{\sum_l S_l}
\end{equation}

\begin{figure}[th]
\begin{center}
\includegraphics[scale=0.7]{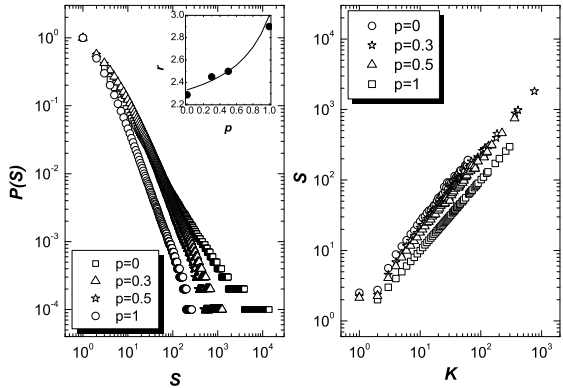}
\caption{Left: Probability distribution $P(S)$. Data are consistent
with a power-law behavior $s^{-r}$. The data are averaged by 20
networks of size $N=10000$ and $m=2$. In the inset, we give the
value of $r$ obtained by data fitting together with the analytical
expression $r=2+1/(3-2p)$. Right: Correlations between strength and
degree. The slopes are all around $1$ indicating linear
relationship. \label{strength}}
\end{center}
\end{figure}

The equation (\ref{strengthequation}) can be rewritten if we notice
the sum of strength at large time $t$ is $\sum_l
S_l\approx2m(p+(1-p)2)t$ and the initial condition $S_i(i)=m,
K_i(i)=m$. So we obtain:
\begin{equation}
\frac{dS_i}{dt}=\frac{3-2p}{4-2p}\frac{S_i}{t}
\end{equation}

\begin{equation}
S_i=m(\frac{t}{i})^{\frac{3-2p}{4-2p}},
K_i=m\frac{S_i+2m(1-p)}{3-2p}\label{f2}
\end{equation}

The strength and degree of vertices are thus related by a linear
equation. It means the model is not only strength-driven but also
degree-driven.

From Eq. (\ref{f2}) we get the power-law distribution of the degree
and strength $P(K)\sim K^{-r}, P(S)\sim S^{-r}$ with the exponent:
$r=2+1/(3-2p)=(7-4p)/(3-2p)$. Obviously $r$ is between 2.33 and 3
when $p$ ranges from $0$ to $1$. When $p=1$, passenger always choose
connect directly, or in other words there are no transfer behavior.
The model is reduced to strength-driven AK model. If we set
$1-p=\delta$, the exponent becomes $r=(3+4\delta)/(1+2\delta)$ which
reproduces the networks generated by the BBV model. It worths
noticing because it indicates that a model based on human behavior
can lead to networks produced by automatic topology and weight
coupling. $\delta$ in BBV model represents the weight increase in
every time period while $1-p$ represents the transfer behavior. The
above finding indicates that transfer behavior can induce weight
dynamics in BBV model.

\begin{figure}[th]
\begin{center}
\includegraphics[scale=0.7]{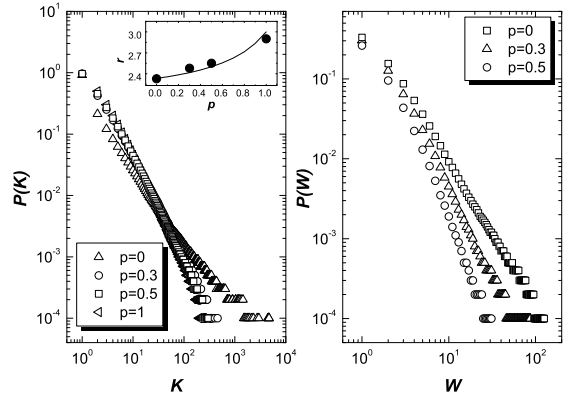}
\caption{Left: Probability distribution $P(k)$. Data are consistent
with a power-law behavior $k^{-r}$. In the inset, we give the value
of $r$ obtained by data fitting together with the analytical
expression $r=2+1/(3-2p)$. Right: Probability distribution $P(w)$.
Data are consistent with a power-law behavior $w^{-a}$. We don't
plot the curve for $p=1$ because in that case all weights are 1 and
the distribution cannot be displayed with an infinite exponent. The
data are averaged by 20 networks of size $N=10000$ and $m=2$.
\label{degree}}
\end{center}
\end{figure}

The time evolution of the weights $w_{ij}$ can also be computed
analytically. Weight $w_{ij}$ between already present node $i$ and
$j$ evolves each time either $j$ or $i$ are selected as transfer
node to another one. The evolution equation can be written as :

\begin{equation}
\frac{dw_{ij}}{dt}=m(1-p)\left(\frac{S_i}{\sum_l
S_l}\frac{w_{ij}}{S_i}+\frac{S_j}{\sum_l
S_l}\frac{w_{ij}}{S_j}\right)
\end{equation}

Noticing the initial condition $w_{ij}(t_{ij})=1$ and $\sum_l
S_l\approx2m(p+(1-p)2)t$, we get

\begin{equation}
w_{ij}=\left(\frac{t}{t_{ij}}\right)^{\frac{1-p}{2-p}}
\end{equation}

Thus the weights also have power-law distribution: $P(w)\sim w^{-a}$
with the parameter $a=(3-2p)/(1-p)$. When $p=0$, $a=3$. When the
$p\rightarrow 1$, $a\rightarrow\infty$. Actually when $p=1$ there
are no transfer behavior and no weight dynamics. All weights are 1.
So the probability distribution has an infinite parameter.

In order to check the analytical predictions we performed numerical
simulations of networks generated by using the present model with
different probability $p$ and network size $N$. The simulation
recovers the analytical results. The probability distribution of
strength, degree and weight are in good agreement with theoretical
predictions. And the strength and degree have linear correlations.
When we fit the data, we use the maximum likelihood method to
estimate the parameter $r$ \cite{8}. See Fig (\ref{strength}) and
Fig (\ref{degree}).

Next we investigate the structural organization of the networks
generated by our model by studying correlations of vertices. Using
the formula defined by Eq. (4) and Eq. (6) of Ref \cite{4}, we
calculate degree-dependent cluster coefficient $C(k)$ and average
nearest-neighbor degree $K_{nn}(k)$ (also called degree-degree
correlations). They both show disassortative behavior with $C(k)$
and $K_{nn}(k)$ decreasing with $k$, which indicates there are
hierarchal structures in the networks. The properties are depending
on the parameter $p$. For large $p$, $C(k)$ and $K_{nn}(k)$ are
quite flat. While the $p$ decreases, the curves grows.

\begin{figure}[th]
\begin{center}
\includegraphics[scale=0.7]{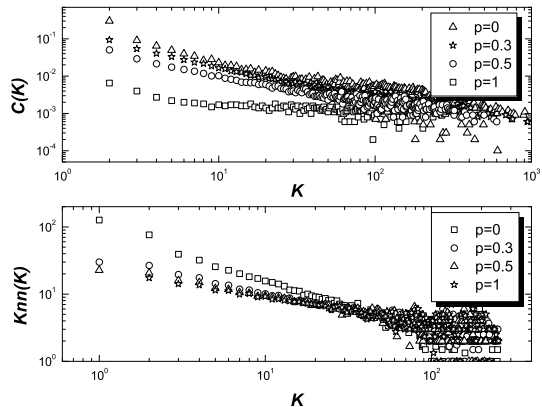}
\caption{Top: Clustering coefficient $C(k)$ depending on the
parameter $p$. Bottom: Average nearest-neighbor degree $K_{nn}(k)$
depending on the parameter $p$. The data are averaged by 20 networks
of size $N=10000$ and $m=2$. \label{clust_coeff}}
\end{center}
\end{figure}

This disassortative behavior can be understood in the dynamical
growth process. Vertices with large connectivities and strengths are
the ones that enter the system early. New vertices are attracted to
preexisting vertices with large strengths. In direct connection
mode, there builds up a disassortative relation between "old" vertex
with high strength and connectivities and "young" vertices with
small connectivities. In transfer mode, the transfer vertices also
have high connections because it generally is the most weighted in
the neighbor of the destination. So there also builds up a
disassortative relation between transfer vertices and the new
vertices. The average nearest-neighbor degree will be higher than
that in direct connection because the transfer mode produces more
highly connected vertices than the direct connection mode. The
cluster coefficient in transfer mode is larger than that in direct
connection mode because transfer mode connects three vertices in one
time period thus produces more correlations between vertices. With
small $p$ there are more transfer behavior in the networks leading
to large $C(k)$ and $K_{nn}(k)$.

In summary we have presented a model for weighted networks evolution
that considers human behavior in addition to weight dynamics and
topology growth. We investigated the evolution of degree, strength
and weight. They are all distributed according to power laws with
exponents dependent on the probability $p$ which determines the
behavior of passengers. Clustering and correlations between vertices
show clear disassortative behavior. The most difference between our
model and previous models is that human behavior is incorporated to
induce weight dynamics instead of automatic weight dynamics. This
reveals the underlying driving force in the growth of transportation
networks and social networks. We believe that the model might
provide a starting point for the realistic modeling that
incorporates human behavior into technological networks modeling.

\section*{Acknowledgments}

The work was supported by Natural Science Foundation of China (NSFC
70432001).

\end{document}